\documentclass{emulateapj}

\slugcomment{To appear in ApJ}

\shorttitle{Raman Scattered O~VI 6825 in V1016 Cyg and HM Sge}
\shortauthors{Lee and Kang} 

\begin{document}

\title{Raman Scattered O~VI~$\lambda$~6825 and the Accretion Disk 
Emission Model in the Symbiotic Stars V1016~Cygni and HM Sagittae}

\author{Hee-Won Lee and Suna Kang} 
\affil{Department of Astronomy and Space Science, 
Astrophysical Research Center for the Structure and Evolution 
of the Cosmos, Sejong University, Seoul, 143-747, Korea}

\begin{abstract}

We present the high resolution spectra of the D type symbiotic stars
V1016~Cygni and HM~Sagittae obtained with the Bohyunsan Optical 
Echelle Spectrograph (BOES), and  investigate 
the double-peaked asymmetric profiles of the Raman scattered O~VI 6825.
This feature is formed through Raman scattering of O~VI 1032
by atomic hydrogen with small scattering cross section of 
$\sim 10^{-23}{\rm\ cm^2}$, requiring a specific condition of coexistence
of a highly ionized emission nebula and a thick neutral region.
By adopting a wind accretion disk model, 
we assume that the O~VI emission region is described by a Keplerian 
thin disk. The Raman scattering occurs in a neutral region near the giant, 
taking in the form of a slow stellar wind, part of which
is ionized by the strong UV radiation from the hot white dwarf.
Using a Monte Carlo technique, we compute the line profiles that are 
modulated by the slow spherical stellar wind from the giant component 
with the ionization front approximated by a hyperboloid.
In order to account for the asymmetry and the existence of a central dip
in the profiles, we add an O~VI resonance scattering region
between the hot white dwarf and the giant star which hinders the incidence
of slightly blue O~VI photons upon the H~I region.
Overall good fits to the observed data are obtained from our model,
which lends support to the accretion disk emission model in these objects.
The best fitting parameters for V1016~Cyg are $v_o=30{\rm\ km\ s^{-1}}$,
 $v_\infty=11{\rm\ km\ s^{-1}}$, and $v_{c}=10{\rm\ km\ s^{-1}}$,
where $v_o$, $v_\infty$ and $v_{c}$ are the velocity of the outer disk
rim, the terminal velocity of the giant wind, and the velocity component
of the resonance scattering O~VI region along the binary axis, 
respectively.  Similar fitting parameters
$v_o=27{\rm\ km\ s^{-1}}$, $v_\infty=10{\rm\ km\ s^{-1}}$ and 
$v_{c}=9{\rm\ km\ s^{-1}}$ are obtained for HM~Sge.
We also investigate the effect of a hot spot in a disk that is well known in
accretion disks in cataclysmic variables. 
However, the introduction of a hot spot in our Keplerian disk
model failed to improve the overall profile fitting quality significantly.  
Brief discussions about our profile analysis in relation to 
bipolar morphology and accretion processes are presented.
\end{abstract}


\section{Introduction}

Symbiotic stars are spectroscopically characterized by prominent emission lines
with TiO absorption bands that are typical of a giant.  They are generally 
known to be wide binary systems consisting of a hot white dwarf and a 
giant star (e.g. Kenyon 1986).  They are usually divided into
'S' and 'D' type symbiotic stars, where 'D' type systems exhibit infrared
excess indicative of warm dust with temperature $T\sim 10^3{\rm\ K}$. 
Many 'D' type symbiotic stars are believed to contain a Mira type
giant as cool component. Their orbital separation of the hot and
cool components may lie in the range of 10-100 AU, from which we expect
the orbital periods of order $10^2{\rm\ years}$ (Whitelock 1987).

The giant component may lose mass
in the form of slow stellar wind, a significant fraction of which can be 
photoionized by the hot white dwarf component.  Taylor
\& Seaquist (1984) investigated the ionization structures in order to
interpret their radio observations of symbiotic stars.  
Many prominent emission lines in symbiotic stars can 
be excellent diagnostic tools to study the physical conditions 
around the white dwarf component (e.g. M\"urset et al. 1991). 

It has been proposed by a number of researchers that some fraction 
of wind material from the giant component may be gravitationally captured 
by the hot component (e.g. Sokoloski
Bildsten \& Ho 2001).  The mass loss rate from the giant component
may range from $10^{-4}-10^{-7}{\rm\ M_\odot\ yr^{-1}}$ depending
on the stellar evolution stage (e.g. Iben \& Tutukov 1996, Luthardt 1992).
Wind accretion in binary star systems was investigated
using the smoothed particle hydrodynamics (SPH) by Theuns
and Jorrissen (1993), who concluded that an accretion dis can be 
formed in their isothermal flow models.
Mastrodemos \& Morris (1998) also adopted an SPH method
to show that an accretion disk can be formed around the white dwarf
component.  There are only limited studies about 
the existence of accretion disks and the insufficient numerical resolution
prevents one from deducing the physical properties of an accretion disk 
in symbiotic stars.  Symbiotic stars being considered 
as a candidate for type Ia supernova progenitors, the mass transfer
process in symbiotic stars can also be a very important subject in other
branches of astronomy.

About a half of symbiotic stars are known to exhibit very broad and
mysterious emission features around 6825 \AA\ and 7088 \AA\ that were 
suspected to originate from highly ionized species (Allen 1980). 
These features were finally identified by Schmid (1989), who proposed the 
Raman scattering origin. According to him, the Raman 
scattering process starts with an O~VI 1032 line photon incident 
upon a hydrogen atom in the ground state, which can be converted 
into an optical 6825 photon 
as a result of an inelastic scattering off the hydrogen atom 
that is left in its excited $2s$ state.  From an incident O~VI 1038 
line photon, we obtain a Raman photon at around 7088 \AA.
The Raman scattering nature of these features is strongly supported by
many observational studies including spectropolarimetric observations by 
Harries \& Howarth (1996) and
contemporaneous UV and optical observations by Birriel et al. (2000).

The wavelength $\lambda_o$ of the inelastically 
scattered radiation is related with the incident wavelength $\lambda_i$ 
and  $\lambda_{Ly\alpha}$ of hydrogen Ly$\alpha$
by
\begin{equation}
{\lambda_o}^{-1}={\lambda_i}^{-1}-\lambda_\alpha^{-1}.
\end{equation}
Differentiation of this equation leads to
\begin{equation}
{\Delta\lambda_o\over\lambda_o}=
\left({\lambda_o\over\lambda_i}\right){\Delta\lambda_i
\over\lambda_i}.
\label{wavelength}
\end{equation}

In the case of Raman scattering involving O~VI 1032, $\lambda_o/\lambda_i
=6.6$.
This provides two important characteristics in the profiles of Raman 
scattered O~VI 6825
that they are broader than their parent emission lines by a factor 
of $\lambda_o/\lambda_i$ and that they are mainly determined 
by the relative motion between the neutral scatterers and line photon 
emitters. Therefore, the profiles are little affected by the direction of the
observer's line of sight.  This is due to the inelasticity 
of Raman scattering and noted to be one of the most important spectroscopic
properties retained by Raman scattered features (e.g. Nussbaumer, Schmid \&
Vogel 1989, Schmid 1989). In this respect, Raman scattered lines provide 
a very interesting opportunity to study the accretion process involving
a slow stellar wind, which is quite unique to symbiotic stars.

Thus far, Raman scattered lines
are found for O~VI 1032, 1038 resonance doublets and He II emission lines
near H~I Lyman series in many symbiotic stars and in a few young
planetary nebulae (e.g. van Groningen 1993, 
P\'equignot et al. 1997, Groves et al. 2002, Lee et al. 2003, Birriel 2004, 
Zhang et al. 2005, Lee et al. 2006b). Raman scattered O~VI 6825 from O~VI 1032
is particularly strong in a number of symbiotic stars, for which a
refined profile analysis is feasible. Furthermore, 
Raman scattered features consist of purely scattered photons, not mixed
with the direct emission, which makes them characterized by strong linear
polarization.

Fairly intensive spectropolarimetric studies of Raman scattered O~VI lines
were carried out by Harries \& Howarth (1996). They found that 
the Raman scattered O~VI 6825 and 7088 show double or
triple peak profiles and that they are strongly polarized with
complicated structures.
Schmid (1996) used a Monte Carlo method to calculate
the profiles and polarization of these scattered features formed
in a slowly expanding neutral region that mimics the stellar wind around
the giant component. A similar study with a slightly different velocity
law of the slow stellar wind was performed by Lee \& Lee (1997b). 

In these studies, no kinematic structure is assumed for the emission
region, which is treated effectively as point-like
without any complicated internal structures.
Instead, the stronger red part of Raman scattered O~VI 6825 is
mainly attributed to scattering at the receding part of the
giant stellar wind that has a bigger geometric covering 
of the O~VI emission region than the approaching part does. 
However, in these models, one has to invoke
a rather high velocity scale $\ge 50{\rm\ km\ s^{-1}}$ of the stellar wind 
from the giant in order to explain the observed Raman profiles.

Lee \& Park (1999) adopted an accretion disk emission model and
proposed the origin of  the multiple peak profiles with strong linear
polarization.  In their picture, the accretion flow
can be affected significantly by the binary orbital motion, which results 
in an asymmetric matter distribution around the white dwarf causing
the stronger red part of the Raman features. 
However, their study lacked detailed calculations that can be directly
compared with the observed profiles.

Both V1016~Cyg and HM~Sge are classified as symbiotic novae or slow novae.
V1016~Cyg underwent a slow nova-like outburst in 1965, which may be
interpreted to be a thermonuclear runaway on the surface of the white dwarf
(Fitzgerald et al. 1966). A similar nova-like optical outburst occurred 
in HM Sge in 1975 (Dokuchaeva 1976). These two symbiotic stars are known 
to exhibit fairly strong Raman scattered O~VI 6825 with
characteristic double peak profiles. In particular, the appearance of 
Raman scattered 6825 in HM~Sge was reported by Schmid et al. (2000) 
in their spectra obtained with the William Herschel Telescope (WHT) in 1998.

Both V1016~Cyg and HM~Sge are D-type symbiotic stars, 
containing a Mira type giant as cool component.
Thus far, no detailed information is available about
the orbital parameters of V1016~Cyg and HM~Sge.
Schild \& Schmid (1996) measured the rotation of position 
angle in the polarized Raman 6825 with an estimation of the
orbital period of $\sim 80{\rm\ yr}$ in V1016~Cyg. However, Schmid (1998)
noted that this estimate may be severely affected by the change of the nebular
structure in this system and the orbital period may exceed 100 years
(e.g. Schmid 2001).  Using the {\it Hubble Space Telescope (HST)} data, 
Brocksopp et al. (2002) 
measured the angular separation of $42.4{\rm\ mas}$ between the white dwarf 
and the Mira of V1016~Cyg, proposing the physical binary separation of
$\sim 84{\rm\ AU}$ with their adopted distance of $2{\rm\ kpc}$.
According to the {\it HST} study of HM~Sge by Eyres et al. (2001), 
the separation of the Mira
and the white dwarf is $\sim 50{\rm\ AU}$ with the adopted distance of
$\sim 1.2 {\rm\ kpc}$.

In this paper, we present our spectra 
of these Raman features in the symbiotic stars V1016~Cyg and HM Sge
secured with  the Bohyunsan Optical 
Echelle Spectrograph (BOES),
and make quantitative comparisons with the profiles 
computed by a Monte Carlo technique adopting an accretion disk emission
model.

\section{Observation and Data}

We observed V1016~Cyg and HM~Sge on the nights of 2005 November 7 through 9 
with the Bohyunsan Optical Echelle Spectrograph installed on the 
1.8~m telescope at Mt. Bohyun. The detector was a 2k$\times$4k
E2V CCD with pixel size $15{\rm\ \mu m}\times 15 {\rm\ \mu m}$, in which
about 80 spectral orders covering $\sim 3700{\rm\ \AA}$ to
$\sim 10,000{\rm\ \AA}$ are recorded in a single exposure.
The 300~$\mu$ optical fiber was used 
to yield the spectroscopic resolving power
of $R=30,000$ with an aperture field of 6.4 arcseconds. 
A more detailed description of BOES can be found in
Lee et al. (2006b) and in Kim et al. (2002). 
The exposure times are from 7000~s and 3600~s for V1016~Cyg and HM~Sge, 
respectively.  The data have been
reduced following standard procedures using the IRAF packages.

Parts of our spectrum around 6825~\AA\ of V1016~Cyg and HM Sge are 
shown in Fig.~1.
In the figure, the line flux is normalized by the local continuum
in order to focus on the profile shape in the current work.
The two profiles are overall similar to each other.
The 6825 feature of HM~Sge is much weaker and therefore of poorer quality
than that of V1016~Cyg, which is indicated by the larger scatter 
in the data.  In both systems, the red part is relatively stronger 
than the blue part, which is noted by many previous researchers 
(e.g. Harries \& Howarth 1996, Schmid et al. 1999).

There exists a narrow emission line at around 6820 \AA\ in HM~Sge.
The presence of this narrow emission feature 
in HM~Sge was also noted in the spectropolarimetric studies by 
Schmid et al. (2000). They concluded that this emission line is
not related with Raman scattering on the ground that it is unpolarized
and narrow. We ignore this feature in the subsequent analysis, by
subtracting it by a single Gaussian $f(\lambda)=f_0 \exp[(\lambda
-\lambda_c)^2/\Delta\lambda^2]$ with $f_0=0.43,\  
\lambda_c=6820.09 {\rm\ \AA}$ and $\Delta\lambda = 0.46{\rm\ \AA}$.

In the case of V1016~Cyg, the local maxima are found at
6823.5 \AA\ and 6830.8 \AA. The separation of these two peaks
corresponds to $48{\rm\ km\ s^{-1}}$ in the velocity space 
of O~VI 1032 emitters. 
The FWZI (full width at zero intensities) is $166{\rm\ km\ s^{-1}}$.
It is noted that there is an asymmetrically extended red wing
that may end at around 6842.7 \AA.

The peaks for HM~Sge are located at 6823.3 \AA\ and 6829.4 \AA,
for which the separation in velocity space of O~VI is $41{\rm\ km\ s^{-1}}$.
This is significantly smaller than that for V1016~Cyg.
The FWZI is $161{\rm\ km\ s^{-1}}$, which is comparable to that
of V1016~Cyg. In contrast with V1016~Cyg, it appears that there is
no (or significantly weak) extended red wing in HM~Sge. However, 
this should be taken with the
caution that the data quality of HM~Sge is much poorer than V1016~Cyg. 

The relative kinematics between 
the O~VI emission region and neutral H~I scattering region in both
V1016~Cyg and HM~Sge is
characterized by velocity scales ranging from 
$20{\rm\ km\ s^{-1}}$ to $80{\rm\ km\ s^{-1}}$. If the FWZI of the
profiles is attributed to the kinematics of the giant wind as
in the models of Schmid (1996) and Lee \& Lee (1997b), then the wind terminal
speed of $80{\rm\ km\ s^{-1}}$ should be required, which is much faster
than the escape velocity of a giant star.

\begin{figure}
\includegraphics[angle=0,scale=.60]{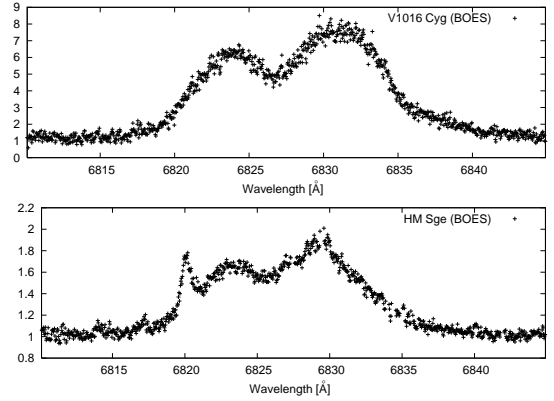}
\caption{Parts of spectra around 6825~\AA\ of the symbiotic 
stars V1016~Cyg and 
HM~Sge obtained with the Bohyunsan Optical Echelle Spectrograph (BOES).
The vertical scale is normalized by the local continuum.
The peak separation is $\sim 48{\rm\ km\ s^{-1}}$ for V1016~Cyg
and $\sim 41 {\rm\ km\ s^{-1}}$ for HM~Sge. The full widths at zero
intensities are $166{\rm\ km\ s^{-1}}$ and $161{\rm\ km\ s^{-1}}$
for V1016~Cyg and HM~Sge, respectively.
}
\end{figure}

\section{Monte Carlo Line Profile Analysis}
\subsection{Accretion Disk Model}

In this work, we assume that the wind accretion occurs in both V1016~Cyg
and HM Sge to form an accretion disk around each white dwarf component.  
According to Model 3 of the SPH studies by Mastrodemos \& Morris (1998),
an accretion disk with size ranging 0.4 - 0.8 AU is formed
in a binary system containing a white dwarf, where the binary orbital
period is 18 years.  We consider that the O~VI emission region is limited
to a part of the whole accretion flow, which is approximated to be Keplerian. 
For simplicity, we assume that
the O~VI emission region is confined to an annular region 
characterized by the inner radius $R_i$
and the outer radius $R_o$.

In this work we assume that each white dwarf component has a mass
of $0.7 {\rm\ M_\odot}$. Noting that the Raman scattered O~VI
6825 in both V1016 Cyg and HM Sge extends about $\Delta v\sim 160{\rm\ km\ 
s^{-1}}$, we infer that the O~VI emission region extends from $\sim 0.05$ AU
to less than 1 AU. 
 

With these ingredients we prepare a characteristic double peak profile
of O~VI 1032, we first consider the disk component with the emissivity
$\epsilon({\bf r})$ that depends only on $r$ through a power law
\begin{equation}
\epsilon({\bf r}) = Ar^{\alpha}, \ R_i<r<R_o
\label{eqalpha}
\end{equation}
where $A$ is a constant.  This functional dependency of emissivity is
achieved in our Monte Carlo code by specifying the location $r$ 
of the initial photon from a uniform random number $r_p$ in the interval
$(0,1)$ with the prescription
\begin{equation}
{r\over R_o} =
\left[\left( {R_i\over R_o} \right)^{\alpha+2}
+\left( 1-\left\{ R_i\over R_o \right\}^{\alpha+2} \right)r_p
\right]^{1/(\alpha+2)}.
\end{equation}

With another random number $r_\phi$ in the interval $(0, 2\pi)$, we
locate the point of generation of the incident photon $P(r,\phi=2\pi
r_\phi)$. Then the photon is assigned a Doppler factor $DF$ along
the propagation direction
\begin{equation}
DF = {\bf k}_i\cdot {\bf v}(P)/c,
\end{equation}
where ${\bf v}(P)=\sqrt{GM_{WD}\over r}(-\sin\phi{\bf\hat
  x}+\cos\phi{\bf\hat y})$ is the local Keplerian velocity at $P$.

According to Eq.~(\ref{wavelength}), the observed wavelength of Raman 
scattered O~VI is determined
by the relative motion between the O~VI emitter and the H~I scatterer
and quite insensitive to the observer's line of sight. 
Furthermore, the orbital parameters of V1016~Cyg and HM~Sge
are only poorly known.
In view of these facts, in our Monte Carlo code, we collect all the photons
scattered inelastically off hydrogen atoms around the giant
component. However, a more refined study should include the effect
of the observer's line of sight with full care.

\subsection{Velocity Modulations from a Slow Stellar Wind}

The direct O~VI 1032 line photons are scattered in a neutral scattering
region, which may be expanding with a velocity $\sim 10{\rm\ km\
  s^{-1}}$. Schmid (1996) investigated the line profile formation of
Raman scattered O~VI in an expanding slow stellar wind using a Monte
Carlo technique (see also Harries \& Howarth 1997). 
A similar study was also conducted by Lee \& Lee
(1997b) using a Monte Carlo method with a slight different kinematical
velocity law for the slow stellar wind.
 
Adopting also a Monte Carlo technique, we extend the work of Lee \&
Lee (1997b) to incorporate the accretion
disk emission model considered in the previous section.  The velocity
of the slow and spherical stellar wind ${\bf v}_H({\bf r})$ at the position 
${\bf r}$ is given by
\begin{equation}
{\bf v}_H({\bf r}) =v_\infty (1-R_*/r)^\beta {\bf\hat r},
\label{eqgwind}
\end{equation}
where $v_\infty$ is the terminal wind velocity and the origin of the
coordinate system coincides with the center of the giant. For
simplicity of the current investigation, we set the power $\beta=1$ of the
velocity law. We introduce a new dimensionless vector defined by
\begin{equation}
{\vec \rho}={\bf r}/R_*.
\end{equation}
Assuming a constant mass loss rate $\dot M$, the density
$n{\bf r})$ is described by
\begin{equation}
n({\bf r}) = n_0\rho^{-2}(1-\rho^{-1})^{-1},
\label{nr}
\end{equation}
where $n_0=\dot M/4\pi\mu R_*^2 m_pv_\infty$ with $m_p$ and $\mu$ 
being the mass of a proton and the mean molecular weight, 
respectively.

In the Monte Carlo calculation, it is essential to locate the next
scattering position ${\vec\rho}_2$ for a given starting 
position ${\vec\rho}_1$ in the direction ${\bf\hat k}_i$, the unit 
wavevector of the photon under consideration.  
In order to accomplish this, an optical depth $\tau_{12}$ is
generated using a uniform random number $r_{tau}$ between 0 and 1 by
\begin{equation}
\tau_{12}=-\ln r_{tau}.
\end{equation} 
In terms of the physical distance $s_{12}$ corresponding to $\tau_{12}$
from the starting position ${\vec\rho}_1$ along the direction
${\bf\hat k}_i$, the new scattering position is expressed as
\begin{equation}
{\vec\rho}_2 = {\vec\rho}_1 + s_{12}{\bf\hat k}_i.
\label{s12}
\end{equation}
The new scattering position $\vec\rho_2$ is related with $\tau_{12}$ by
\begin{equation}
\tau_{12} = \int_{{\vec\rho}_1}^{{\vec\rho_2}}\ n({\bf r}) 
\sigma_{tot}\ dl,
\label{tau12d}
\end{equation} 
where $\sigma_{tot}$ is the sum of the cross sections for Rayleigh and
Raman scattering. Lee \& Lee (1997a) proposed the value 
$\sigma_{tot}=42\sigma_T=2.8\times 10^{-23}{\rm\ cm^2}$
with $\sigma_T=6.6\times 10^{-25}{\rm\ cm^2}$ being the Thomson
scattering cross section, which we adopt in this work. 

We introduce the representative optical depth 
$\tau_0$ defined by
\begin{equation}
\tau_0 = n_0 R_*\sigma_{tot} =8.8 \dot M_{-6}\ R_{*13}\ v_{\infty10},
\end{equation}
where $\dot M_{-6}=\dot M/(10^{-6}{\rm\ M_\odot\ yr^{-1}})$,
$R_{*13}=R_*/(10^{13}{\rm\ cm})$ and $v_{\infty10}=v_{\infty}/(10{\rm\
  km\ s^{-1}})$. 
The exact choice of $\tau_0$ and $\sigma_{tot}$ 
affects the ionization
structure and hence the exact profile shape. However, the ionization structure
is also dependent on the UV radiation from the hot component. In view of this,
we set $\tau_0=2.5$ in our simulations, in order to reduce the parameter 
space of investigation. 

We also define $b$ as the impact parameter of the photon path with respect to
the giant center, and the parameter $s$ measures the physical distance
along the photon path. For the sake of convenience, we set $s=0$ 
at the foot of the perpendicular from the giant center to the photon
path, so that
\begin{equation}
s=\pm\sqrt{\rho^2 - b^2},
\label{impact}
\end{equation}
where negative $s$ is obtained when ${\bf\hat k}_i\cdot {\vec\rho}_1
<0$. In terms of these parameters, the integral (\ref{tau12d}) can be
rewritten as
\begin{equation}
\tau_{12} =\int_{s_1}^{s_2}  n(s)\ \sigma_{tot}\ ds
=\tau_0\int_{\rho_1}^{\rho_2} {d\rho\over\sqrt{\rho^2-b^2}(\rho-1)}. 
\label{tau12}
\end{equation}
In terms of a new function $u(\rho)$ defined by
\begin{equation}
u(\rho)={1\over\sqrt{|b^2-1|}}[\rho-1+\sqrt{\rho^2-b^2}],
\label{urho}
\end{equation}
the integral in Eq.(\ref{tau12}) can be explicitly expressed as
\begin{equation}
\tau_{12}={2\tau_0\over\sqrt{|b^2-1|}}[\tan^{-1}u(\rho_2)
-\tan^{-1|}u(\rho_1)].
\end{equation} 
Using the inverse relation of Eq.(\ref{urho})
\begin{equation}
\rho=b\cosh \ln [(\sqrt{b^2-1}u+1)/b]
\end{equation}
this relation is directly inverted to obtain $\rho_{2}$ 
in terms of $\tau_{12}$ and $\rho_1$.

A more complicated situation is obtained when $b>1, 
{\bf\hat k}_i\cdot{\bf r}_1<0$, in which case we consider
$\tau_b$, the optical depth to the point of impact, defined by
\begin{equation}
\tau_b 
= {2\tau_0\over\sqrt{b^2-1}}
\left[\tan^{-1}u_1-\tan^{-1}\left({ b-1 \over b+1 }\right)^{1/2}\right].
\end{equation}
If $\tau_{12}>\tau_b$, we consider
\begin{equation}
\tau_{12}=\tau_b+\int_0^{s_2} n(\rho)\ \sigma_{tot}\ ds,
\end{equation}
from which we obtain
\begin{equation}
u(\rho_2)=\tan{ \sqrt{b^2-1}(\tau_{12}-\tau_b)\over 2\tau_0}.
\end{equation}

With $\rho_2$, $s_{12}$ is determined, and the location of 
the next scattering position ${\vec \rho}_2$ is then given 
by Eq.(\ref{s12}).  With a branching
ratio $b_{Ram}\sim 0.15$, the scattering process is Raman, in which case
we assume that the Raman photon escapes from the region without further
interaction. If the scattering process is Rayleigh, then the photon
propagates along the new direction ${\bf\hat k}_o$, which is chosen
from the Rayleigh phase function (e.g. Schmid 1996, Lee \& Lee 1997a).

\subsection{Ionization Fronts}
\begin{figure}
\plotone{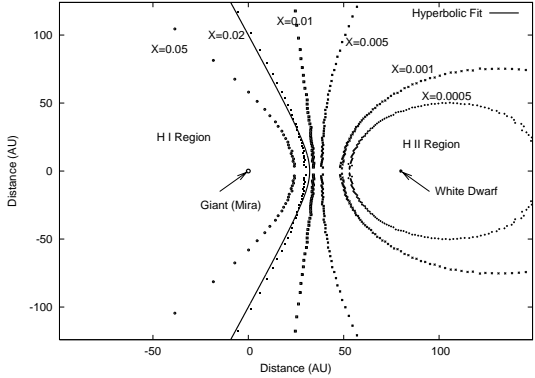}
\caption{Ionization structure in a symbiotic star 
for a range of ionizing luminosities
represented by the parameter $X$ defined in Eq.~(\ref{ionx}).  
The dots at the origin and 
at the point $(80,0)$ stand for a mass losing giant and a hot white
dwarf.  The volume of the H~I region increases as $X$ decreases.
The ionization front corresponding to $X=0.02$ is approximated by
a hyperbola given by $y^2=6.5(x-40)^2-400$.}
\label{ifront}
\end{figure}

The part of the stellar wind from the giant facing the hot white
dwarf will be ionized by strong far UV radiation. Taylor \& Seaquist
(1984) investigated the ionization structures that
depend the density and the ionizing luminosity. They introduced a
a parameter $X$ defined by
\begin{equation}
X = {4\pi\mu^2m_p^2\over\alpha} aL_{ph} (v_\infty/\dot M)^2.
\label{ionx}
\end{equation}
Here, $\alpha$ is the recombination coefficient, 
$L_{ph}$ is the luminosity of ionizing radiation, and
$a$ is the photoionization cross section.

Then, the ionization front is described by the equation
\begin{equation}
X=f(u,\theta) \equiv\int_0^u n({\bf r})^2 s^2 ds
\end{equation}
where $s$ is the distance measured from the white dwarf. Taylor \&
Seaquist (1984) provided the closed form of $f(u,\theta)$ 
using the density law $n({\bf r})\propto r^{-2}$, which is slightly
different from the one given by Eq.~(\ref{nr}) adopted in this work.
Although the description of Taylor \& Seaquist (1984) is also
approximately valid for the current work, we present the detailed 
ionization structure in a closed form in this subsection
for the sake of completeness. 

With the density law given by Eq.~(\ref{nr}) the necessary integral to
be carried out is
\begin{equation}
I=\int{s^2 ds\over \rho^2(\rho-1)^2}=
\int{(\rho_i\cos\phi\pm\sqrt{\rho^2-b^2})^2 d\rho
\over \rho(\rho-1)^2 \sqrt{\rho^2-b^2}},
\end{equation}
where $\rho$ and $s$ are related by the impact parameter $b$ as in 
Eq.~(\ref{impact}). 

For $b=\epsilon^{-1}>1$, an explicit expression
of the integral $I$ is given by
\begin{eqnarray}
I &=& \rho_i\cos2\phi\csc\phi\tan^{-1}\sqrt{(\epsilon\rho)^2-1}
+{(\epsilon^2-\csc^2\phi)\sqrt{\epsilon^2\rho^2-1}
\over \epsilon(\rho-1)(1-\epsilon^2)}
\nonumber \\
&+&
{4\epsilon^2\cot^2\phi-2\epsilon^2+2\csc^2\phi \over 
\epsilon(1-\epsilon^2)^{3/2}}
\tan^{-1}\left[ \left({1+\epsilon\over 1-\epsilon}\right)
\left({\epsilon\rho-1\over\epsilon\rho+1}\right) \right]^{1/2}
\nonumber \\
& \pm&
2\rho_i\cos\phi 
\left(\ln{\rho\over\rho-1}
-{1\over\rho-1}\right).
\label{eqifront1}
\end{eqnarray}
A more detailed calculation is given in the appendix.

In Fig.~\ref{ifront}, we show the ionization fronts for various values 
of $X$. The horizontal axis coincides with the line connecting
the two stars. The dot at the origin stand for the mass losing giant.
We place the white dwarf at the point $(80,0)$, considering the 
binary separation $\sim 80{\rm\ AU}$ proposed by Brocksopp et al. (2004).
The ionization front corresponding to $X=0.02$ is fitted by a hyperbola
given by $y^2=6.5(x-40)^2-400$. The angle between the horizontal axis
and an asymptote is $69^\circ$. This is somewhat larger than
the value proposed for V1016~Cyg by Jung \& Lee (2004).  Nevertheless,
we perform Monte Carlo calculations using this value, because the exact
value of the half opening angle affects only the overall flux of the Raman 
6825 feature and the profile shape is less sensitive.

\section{Results}

\subsection{Pure Keplerian Disk}

\begin{figure}
\plotone{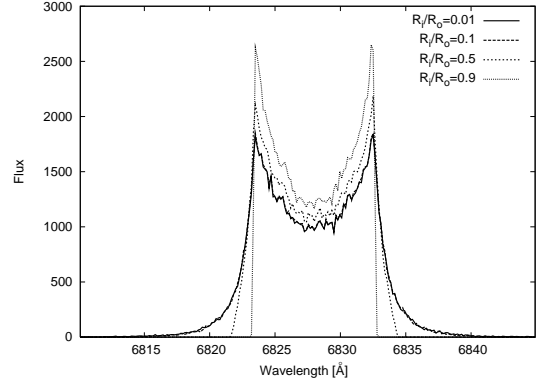}
\caption{
Angle-averaged Monte Carlo line profiles of Raman scattered
6825 from Keplerian disks for various $R_i/R_o$ with $R_i$ and $R_o$
being the inner radius and the outer radius of the disk.
A stationary scattering H~I region is assumed.
Main peaks are found at $\pm v_o$, the velocity
at the outer rim of the disk, which is set to $30{\rm\ km\ s^{-1}}$.
}
\label{purekep}
\end{figure}

In Fig.~\ref{purekep}, we show the profiles of Raman scattered 6825
formed in a stationary H~I region relative to a Keplerian disk.
The H~I region is modeled as the hyperboloid given in Fig.~\ref{ifront}
of which the asymptotic cone has an half opening angle $\theta_o=69^\circ$.  
In the figure, we fix the velocity $v_o=30{\rm\ km\ s^{-1}}$
at the outer edge $r=R_o$, and varied the inner radius $R_i$ from
$0.01R_o$ to $0.9R_o$.  In the absence of a bright spot, the emission
profile is symmetric with respect to the line center.

It is quite clear that the main peaks correspond to the velocity $v_o$
at the outer rim of the disk, which dominantly contributes to the
whole line flux. The emission region is almost ring-like in
the case $R_i=0.9R_o$, where the profile has a sharp edge at $\pm v_o$.
With decreasing $R_i$ the profile broadens around the sharp edges
at $\pm v_o$. However, when $R_i \le 0.1R_o$, the profiles are effectively
similar, because the inner disk is simply physically too small 
to contribute to the overall profile significantly. 
In view of this, we fix the ratio of the inner and outer radii to
be $R_i/R_o=0.1$ from now on.

\begin{figure}
\plotone{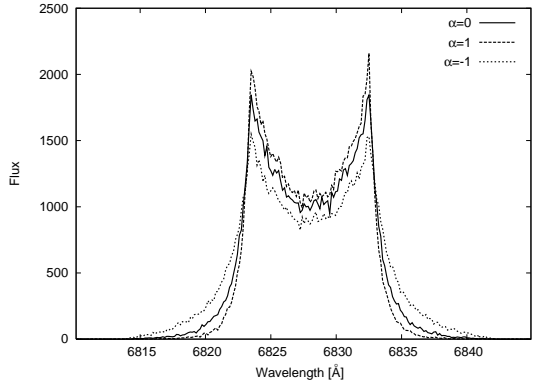}
\caption{Line profiles from a Keplerian disk with the emissivity 
given by $\epsilon(r) = Ar^\alpha$, where $\alpha=\pm1$ and $\alpha=0$. 
Solid line shows a profile for $\alpha=0$, dotted line for $\alpha=-1$
and long dashed line for $\alpha=1$. Far wings,
contributed from the inner disk region, get stronger 
as $\alpha$ decreases.
}
\label{alpha}
\end{figure}

In Fig.~\ref{alpha}, we show three line profiles for $\alpha=-1, 0$ and 1,
where $\alpha$ is defined in Eq.~\ref{eqalpha}. In this figure, $R_i=0.1R_o$
is fixed in order to focus on the dependence of profiles on the functional 
form of the emissivity. With negative $\alpha$, the contribution from
the inner disk region becomes larger, which leads to a broader 
profile with enhanced wings. 
However, the locations of the main peaks are fixed at velocities
$v=\pm v_o$, yielding qualitatively overall similar profiles 
irrespective of the detailed values of $\alpha$.  
From this consideration, we fix $\alpha=0$ in the remainder of this paper.

\subsection{Profile Modulation by the Giant Wind}

\begin{figure}
\plotone{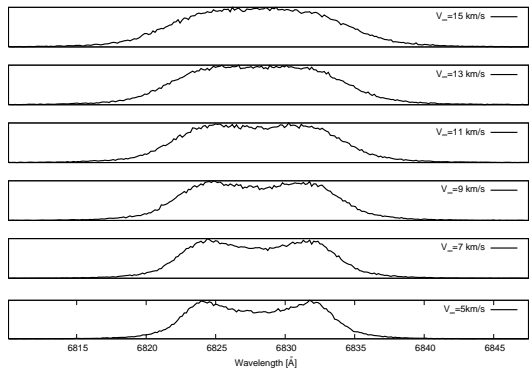}
\caption{Angle-averaged line profiles of Raman scattered 6825 
from a Keplerian disk that are formed in a spherically 
expanding H~I region around the Mira component characterized 
by the terminal wind velocity $v_\infty$.  The H~I region is partially 
ionized with the ionization front given by the hyperboloid 
shown in Fig.~\ref{ifront}.  With smaller $v_\infty$ the double-peak 
structure is more conspicuous, and it is completely erased when
$v_\infty \ge 15{\rm\ km\ s^{-1}}$. 
}
\label{giantwind}
\end{figure}

In Fig.~\ref{giantwind}, O~VI line photons are incident on the
stellar wind around the giant that is expanding according to 
Eq.~(\ref{eqgwind}).
In this figure, the wind terminal velocity $v_\infty$ has a range 
from $5{\rm\ km\ s^{-1}}$ to $15{\rm\ km\ s^{-1}}$. 
Here, we also fix the inner radius $R_i=0.1R_o$, and other parameters
to be the same as those considered in Fig.~\ref{purekep}. Therefore
Fig.~\ref{purekep} is obtained when we set $v_\infty = 0$.

As is seen clearly in the figure, the sharp edges near $v=\pm v_o$
are smoothed by the expanding motion of the H~I region.
When $v_\infty$ exceeds $13{\rm\ km\ s^{-1}}$
the double peak structure can hardly be noticed.  From this, it is
concluded that the giant stellar wind with terminal velocity $<
13{\rm\ km\ s^{-1}}$ is consistent with the double peak profiles
observed in V1016~Cyg and HM~Sge.

However, the profile modulation from other sources such as
turbulence in the emission region may give rise to similar
profiles shown in Fig.~\ref{giantwind}. We do not consider this
possibility in the current work.

\subsection{O~VI Resonance Scattering Region}

\begin{figure}
\plotone{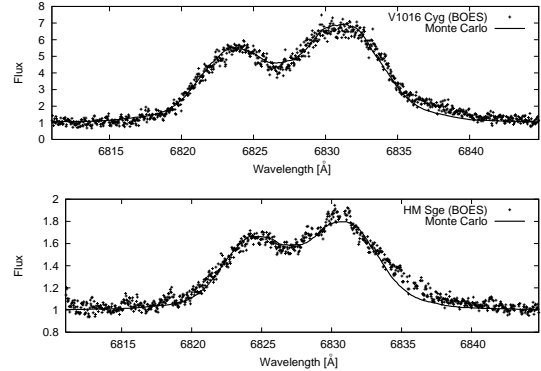}
\caption{Best fitting profiles from our Monte Carlo
calculations and the BOES spectra of V1016~Cyg and HM~Sge. Dots show 
the BOES spectra and solid lines represent our Monte Carlo results.
The fitting parameters for V1016~Cyg are $v_o=30{\rm\ km\ s^{-1}},
v_\infty=11{\rm\ km\ s^{-1}}, v_c=-10{\rm\ km\ s^{-1}}$ and
$\Delta v_c = 9{\rm\ km\ s^{-1}}$.
For HM~Sge they are $v_o=26{\rm\ km\ s^{-1}},
v_\infty=10{\rm\ km\ s^{-1}}, v_c=-7{\rm\ km\ s^{-1}}$ and
$\Delta v_c = 6{\rm\ km\ s^{-1}}$. 
}
\label{bestfit}
\end{figure}

It is clear that the redward asymmetry exhibited in the Raman 6825 features
of V1016~Cyg and HM~Sge can not be accounted for by a pure Keplerian disk
emission model.  It is quite uncertain whether we are seeing an excess in
red emission or a deficit in the blue part, or possibly both. An excess
red emission may correspond to a hot spot in an accretion disk which will
be discussed in the following subsection and in this subsection we consider
the first possibility. 

In order to account for the stronger red part in the observed profiles,
another component need to be added in the previous model considered 
in Fig.~\ref{giantwind}. Therefore, we introduce an O~VI region between
the white dwarf and the giant. This additional O~VI region is assumed
to have a very low density so that the O~VI emission from this region is 
negligible compared to the Keplerian emission disk. Nevertheless, 
the line center optical depth exceeds unity in a velocity space 
centered at $v_c<0$. Schmid et al. (1999) investigated O~VI 1032 profiles
and compared with those of non-resonant O~VI lines to conclude that
the line transfer effect due to resonance scattering of O~VI 1032 
is present in symbiotic stars including V1016~Cyg.

More specifically, we assume that the scattering optical depth
by the hot wind component is given by a Gaussian function 
\begin{equation}
\tau_w = \exp[(v-v_c)^2/\Delta v_c^2].
\label{eqhotwind}
\end{equation}
This prescription hinders the incidence of those line photons
with the Doppler factor near $v_c/c$ upon the scattering region.
However, the velocity modulation due to the giant wind fills
the gap significantly, allowing one to obtain a double-peak profile
with some suppresion in the red part.

Fig.~\ref{bestfit} shows the best fitting profiles 
to out BOES spectra of V1016~Cyg
and HM~Sge. For V1016~Cyg, the velocity $v_o$ at the outer rim
of the accretion disk is $v_o=30{\rm\ km\ s^{-1}}$ and the
terminal velocity of the giant wind $v_\infty = 11{\rm\ km\ s^{-1}}$.
The corresponding values for HM~Sge are
$v_o = 26{\rm\ km\ s^{-1}}, v_\infty = 10{\rm\ km\ s^{-1}}$.  
The parameters adopted for the hot wind component in V1016~Cyg are
$v_c= -10{\rm\ km\ s^{-1}}$ and $\Delta v_c = 9{\rm\ km\ s^{-1}}$
and  $v_c= -7{\rm\ km\ s^{-1}}$ and $\Delta v_c = 6{\rm\ km\ s^{-1}}$
for HM~Sge.

\begin{figure}
\plotone{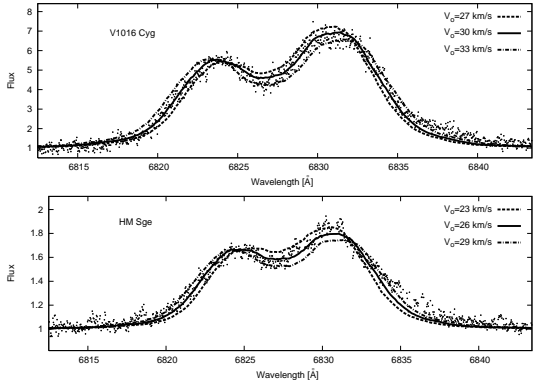}
\caption{Line profiles from our Monte Carlo
calculations for various velocities $v_0$ at the disk outer rim. 
Dots show the BOES spectra and lines represent our Monte 
Carlo results. Solid lines show our best fit profiles shown in 
Fig.~6.
}
\label{fitacc}
\end{figure}

In  Fig.~\ref{fitacc}
we present the line profiles for various velocities $v_o$ at the disk outer
edge ranging from $24{\rm\ km\ s^{-1}}$ to $33{\rm\ km\ s^{-1}}$
with $v_\infty, v_c$ and $\Delta v_c$ fixed to the best fit values
of Fig.~(\ref{bestfit}).
The middle panel shows the best fit line profiles in Fig.~\ref{bestfit}, 
whereas the top
and bottom panels show line profiles with faster and slower disks,
respectively. Considering the much poorer fits shown in the top and
bottom panels, we may conclude that the outer rim velocities are
constrained to be in the range $v_o=30\pm3{\rm\ km\ s^{-1}}$
for V1016~Cyg and $v_o=26\pm 2{\rm\ km\ s^{-1}}$ for HM~Sge.
Adopting a typical white dwarf mass of $M_{WD}=0.7{\rm\ M_\odot}$,
the main O~VI emission region resides from the hot white dwarf component
at $\sim 0.7$ AU for V1016~Cyg and $\sim 0.9$ AU for HM~Sge.

\begin{figure}
\plotone{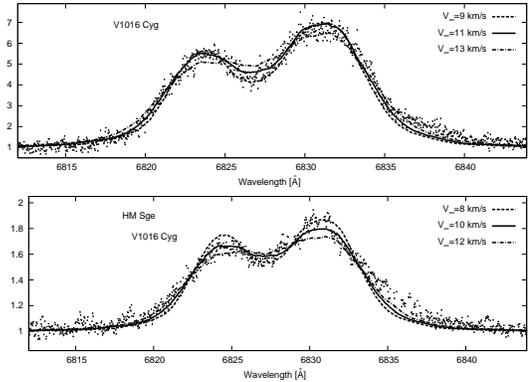}
\caption{Line profiles from our Monte Carlo
calculations for various terminal giant wind velocities $v_\infty$
with $v_o, v_c$ and $\Delta v_c$ fixed to the best fit values of 
Fig.~(\ref{bestfit}).  Dots show the BOES spectra 
and lines represent our Monte Carlo results. The best fit profiles
in Fig. 6 are shown by solid lines. 
Larger values of $v_\infty$ tend to yield better fits at far wing parts
but erase the double peak structure in the main parts.
}
\label{fitwind}
\end{figure}

Fig.~\ref{fitwind} shows the Monte Carlo line profiles
computed for various terminal giant wind velocities $v_\infty$
from $8{\rm\ km\ s^{-1}}$ to $13{\rm\ km\ s^{-1}}$
also with $v_o, v_c$ and $\Delta v_c$ fixed to the best fit
values introduced in Fig.~(\ref{bestfit}). 
Similarly as in Fig.~\ref{fitacc}, the middle panels show our best
fit profiles in Fig.~\ref{bestfit} for comparison. 
As is discussed in the previous section, larger giant wind velocities
tend to erase the double peak structure, whereas overall sharper
profiles result from  smaller wind velocities yielding poor fits at the
outer parts of the observed profiles. 
It is also interesting to note that the wind terminal velocity $v_\infty \sim
10{\rm\ km\ s^{-1}}$ for both V1016~Cyg and HM~Sge 
is quite similar to the escape velocity of a typical giant star.

\subsection{Hot Spots}

\begin{figure}
\plotone{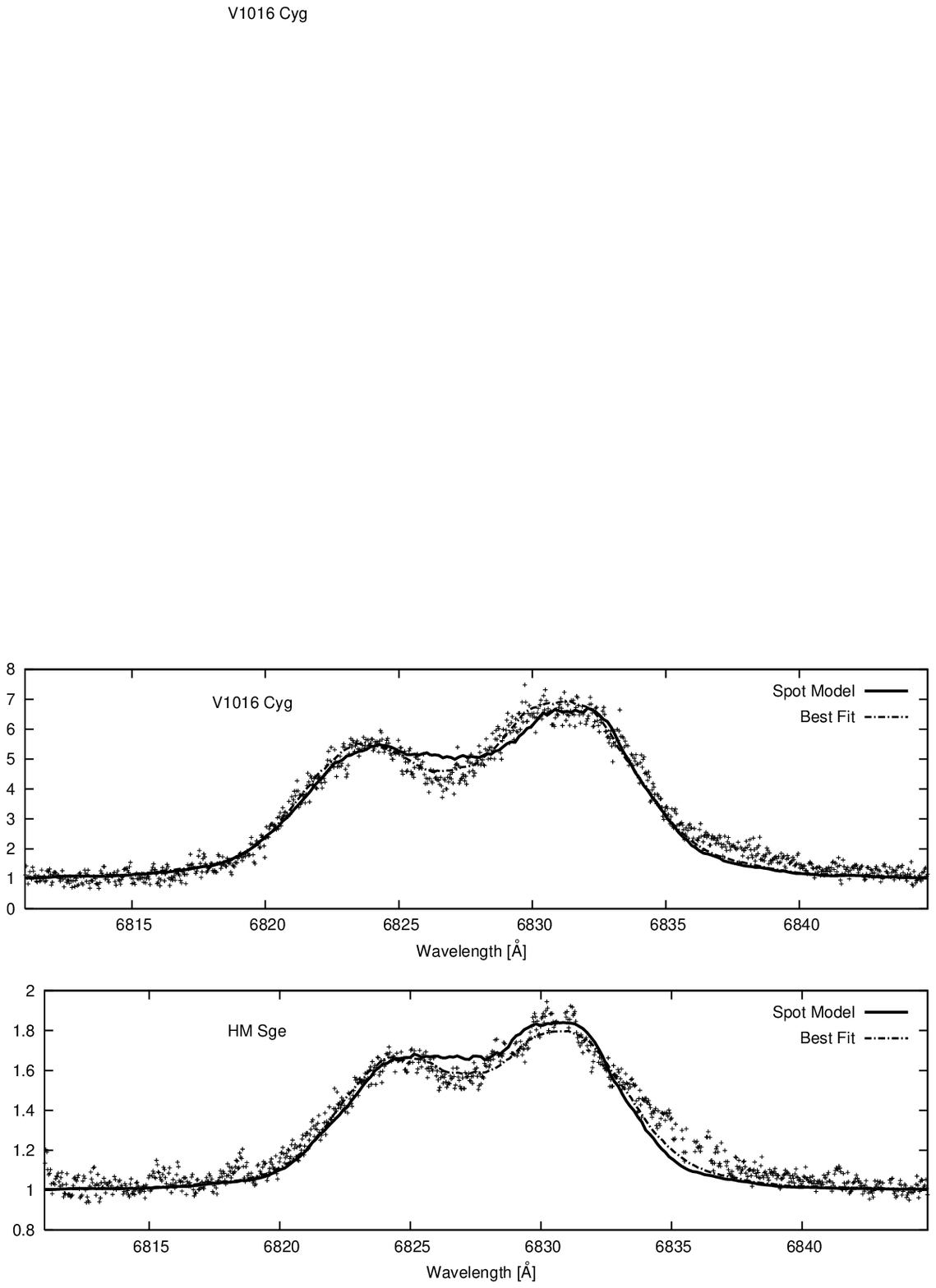}
\caption{Monte Carlo line profiles from a Keplerian disk with a hot spot.
The solid lines show our Monte Carlo data and the dots show the BOES
data. The dotted lines show the best fit profiles given
in Fig.~\ref{bestfit}.  The fitting parameters
are $v_o=33{\rm\ km\ s^{-1}}$ and $v_\infty = 9{\rm\ km\ s^{-1}}$
for V1016~Cyg and  $v_o = 27{\rm\ km\ s^{-1}}$ and $v_\infty=
9{\rm\ km\ s^{-1}}$ for HM~Sge. 
}
\label{hotspot}
\end{figure}

As is well known from spectroscopic studies
of cataclysmic variables, an accretion disk may have a bright spot
which is formed by the impact of the accretion stream near the outer
edge of the accretion disk (e.g. Warner 1995). 
The existence of a similar feature in
symbiotic systems is not clear.  In this subsection, we investigate 
the effect of the introduction of a hot spot in a Keplerian disk 
on the line profile. 

In our Monte Carlo code, photons generated in the
specified region satisfying $r>0.8r_o$ and $|\phi+0.7\pi|<0.1\pi$ is assigned
a statistical weight of 4.8, so that this region is 4.8 times brighter 
than other part of the emission region.  This mimics the bright spots well 
known in cataclysmic variables, for which the
accretion process takes place via Roche lobe overflow. 
Here, we do not consider the hot wind component that provided central
absorption part in Fig.~\ref{bestfit}.

Fig.~\ref{hotspot} shows our profiles. The solid lines show the profiles
from a Keplerian disk with a hot spot, whereas the dots show the BOES
data. For the sake of comparisons, we also show the best fit profiles given
in Fig.~\ref{bestfit} by dotted lines.  The fitting parameters for
V1016~Cyg are $v_o=33{\rm\ km\ s^{-1}}$ and $v_\infty = 9{\rm\ km\ s^{-1}}$.
For HM~Sge, they are $v_o = 27{\rm\ km\ s^{-1}}$ and $v_\infty=
9{\rm\ km\ s^{-1}}$. It should be noted that the disk outer rims move
slightly faster in this model than the ones considered 
in the case of the best fit model.

The profiles provide good fits to the outer parts
$\lambda<6824{\rm\ \AA}$ and $\lambda>6830{\rm\ \AA}$, whereas poor
fits are obtained near center. The introduction of a hot spot enhances 
the red peak. However, the conspicuous central dips in the observed data
are not obtained in pure Keplerian disk models with $v_o >25{\rm\ km\ s^{-1}}$ 
for $v_\infty >8{\rm\ km\ s^{-1}}$. Lowering $v_\infty$ makes the fit worse
by narrowing the widths of the two peaks.

\section{Summary and Discussion}

In this work, we present our profile analysis of Raman scattered
O~VI 6825 in the symbiotic stars V1016~Cyg and HM~Sge, assuming that
O~VI line photons are generated from a thin Keplerian disk with
a scattering region between the white dwarf and the Mira giant. 
The neutral scattering region is assumed to be a spherical 
slow stellar wind with terminal wind velocity $v_\infty$ 
with an ionization front approximated by a hyperboloid.  
The best fitting parameters are $v_o=30{\rm\ km\ s^{-1}}$, $v_\infty=
11{\rm\ km\ s^{-1}}$ and $v_w = -10{\rm\ km\ s^{-1}}$ for V1016~Cyg
and $v_o=26{\rm\ km\ s^{-1}}$, $v_\infty = 10{\rm\ km\ s^{-1}}$
and $v_w=-7{\rm\ km\ s^{-1}}$ for HM~Sge. 

The overall fitting quality is quite insensitive to our choice of
the inner radius of the O~VI emission region and the functional
dependence of emissivity on radius. Furthermore, the profile shape
is little affected by the exact shape
of the ionization structure, which is determined by the mass loss rate
of the Mira component and the ionizing luminosity from the white dwarf
component. However, the ionization structure mainly affects the total 
flux of Raman scattered 6825.  In this point, an independent study 
should be added to current work in order to adequately probe 
the ionization structure and mass loss and transfer processes in
symbiotic stars (e.g. Jung \& Lee 2004).

Although the introduction of a hot spot in our Keplerian disk
model failed to improve the profile fitting quality in the case
of V1016~Cyg and HM~Sge,
this should not mean to exclude the possibility of the asymmetrical
emission in symbiotic stars. As Harries \& Howarth (1996) reported,
triple peak structures in Raman scattered 6825 in symbiotics are 
more prevalent than double peak profiles in symbiotic stars. 
It is quite probable that more complicated profiles may be
fitted using more ingredients in the emission and/or scattering regions.

The resonance scattering O~VI component considered in section 4.3
may be envisioned as a part of ionized giant
stellar wind, in the vicinity of the inner Lagrange point. 
In order to scatter a significant fraction of slightly blue O~VI photons, the
region should move toward the main O~VI line emission region 
around the white dwarf. We may note that the bestfit speed $v_c$ in 
Fig.~\ref{bestfit} is slightly less than the giant wind terminal
velocity. Because $v_c$ is the velocity component along the direction
connecting the giant and the white dwarf, we may interpret that the
resonantly scattering O~VI region passes through the inner 
Lagrange point making an angle
\begin{equation}
\theta_i \simeq \sin^{-1}{0.8} = 50^\circ
\end{equation}
under the assumption that the speed of the region is coincident with 
the wind terminal velocity $v_\infty$.
With the wide binary orbit of $50-80$ AU the orbital speed of
the giant component in both V1016~Cyg and HM~Sge is comparable
to $v_c$.  Although we only consider the spherical wind around
the Mira type giant component in the current work, it is also a 
possibility that the wind around the giant component 
also possesses an azimuthal component, in which case this inference
of $\theta_i$ should be corrected appropriately. Refined hydrodynamical
studies may shed more light on the mass transfer processes in symbiotic
stars.

The red wing excess in V1016~Cyg that is not well fitted by our model
may indicate the existence of another emission component and/or
a neutral scattering region. Spectropolarimetric studies by 
Harries \& Howarth (1996) show that the far red wing part is polarized 
in the direction
perpendicular to that for the remainder part. These facts are consistent
when the additional emission and/or neutral component is receding 
with a velocity of several
tens of ${\rm km\ s^{-1}}$ perpendicular to the binary orbital plane.

It is unclear that this component is associated with the bipolar
structure of these objects. Solf (1983, 1984) detected a bipolar outflow
in [N II] lines in HM~Sge. 
The velocity scale of several tens of km s$^-1$
is much smaller than the usual velocity scale associated with the fast
hot wind that may emanate from a deep inner region of an accretion disk
in the vicinity of the hot star.  Schmid et al. (1999) reported 
the existence of broad wing components in O~VI 1032, 1038 observed with
ORFEUS. Even though the broad wings are consistent with the electron
scattering origin, a fast bipolar wind around the hot component
can also be a candidate.  It is an interesting possibility that 
the component responsible for resonance scattering of O~VI is a clumpy 
one driven by the fast hot wind moving in the polar directions.  
  More interesting results are expected when 
photoionization calculations and hydrodynamical studies can be combined 
with spectropolarimetric polarimetric observations of these symbiotic objects.


It is also interesting that these two symbiotic stars exhibit He II Raman
scattered features blueward of H Balmer lines. In particular, Raman scattered
He II 4850 features of V1016 Cyg and HM Sge exhibit no multiple peak 
structures (e.g. Jung \& Lee 2004, Birriel 2004). 
Because the cross section for He II Raman scattering is larger than
that for O~VI Raman scattering by two orders of magnitude, we may expect
that the scattering region is much more extended than the Raman O~VI
counterpart.

The profiles of Raman O~VI 6825 in V1016~Cyg and
HM~Sge are totally different from other high ionization lines including
He~II. For example, He~II~4686 emission lines of V1016~Cyg and HM~Sge
are single-peaked with slight blue asymmetry.  
Robinson et al. (1994) considered line profiles from accretion disks
in symbiotic stars, and it appears to be difficult to infer the existence
of an accretion disk in symbiotic stars from these emission lines. 
It may be that the He~II emission region is more extended than the O~VI
emission region. It is beyond the scope of this work to model other
emission lines and more sophisticated models including photoionization
and hydrodynamics are required. 

The disparity of the profiles in Raman scattered O~VI 6825 and Raman
scattered O~VI 7088 can be an important clue to the structure of the
accretion disk in symbiotic stars.  A number of researchers
pointed out the relative weakness in the blue part of Raman O~VI 7088
(e.g. Schmid et al. 1999, Harries \& Howarth 1996). The profile 
difference in resonance doublets of $S_{1/2}-P_{1/2,3/2}$ was reported
in symbiotic stars and young planetary nebulae (e.g. Feibelman 1983).

\acknowledgements
We thank the staff at the Bohyunsan Observatory with our particular
gratitude to Kang Min Kim and Byung Cheol Lee. We are also grateful to
Hwankyung Sung for useful discussions on our spectroscopic
observation. We thank an anonymous referee, whose suggestions improved
the presentation of the current paper.  This work is a
result of research activities of the Astrophysical Research 
Center for the Structure and Evolution of the Cosmos (ARCSEC) 
funded by the Korea Science and Engineering Foundation.

\appendix 

\section{The Ionization Fronts}

We describe in detail the photoionization calculation for the ionization
fronts, which was briefly discussed in section 2.

The distance from the giant is measured in units of $R_*$, the radius of the
giant, by $\rho = r/R_*$. Let $\rho_i$ be the separation of the giant
and the white dwarf.  Along a path from the white dwarf making an angle $\phi$
with the line connecting the giant, the physical distance is measured 
by the parameter $s$ satisfying
\begin{equation}
s =
\cases{\rho_i\cos\phi - \sqrt{\rho^2 -b^2} \quad {\rm for}\
s\le \rho_i\cos\phi \cr
\rho_i\cos\phi + \sqrt{\rho^2 -b^2} \quad {\rm for}\
s\ge \rho_i\cos\phi \cr
}
\end{equation}
where $b=\rho_i\sin\phi$ is the impact parameter from the giant to the path of 
the ionization radiation from the white dwarf.

With the density law
\begin{equation}
n(\rho) = {n_0\over \rho(\rho-1)},
\end{equation}
where the distance $r$ from the giant center is measured in units of
$R_*$ by the parameter $\rho=r/R_*$.

The recombination rate along the path of ionization radiation
originating from the white dwarf is proportional to the density squared
along the path, given by
\begin{equation}
I = \int n(\rho)^2 s^2 ds.
\label{int1}
\end{equation}
In terms of $\rho$, this integral can be re-written as
\begin{eqnarray}
I &=& \int{(\rho_i\cos\phi\pm\sqrt{\rho^2-b^2})^2\over \rho(\rho-1)^2
\sqrt{\rho^2-b^2}} d\rho 
\nonumber \\
&=&\rho_i^2\cos^2\phi I_1+I_2\pm 2\rho_i I_3,
\end{eqnarray}
We define the three integrals as
\begin{eqnarray}
I_1 &=& \int{d\rho\over \rho(\rho-1)^2\sqrt{\rho^2-b^2}} 
\nonumber \\
I_2 &=& \int{\sqrt{\rho^2-b^2}d\rho\over \rho(\rho-1)^2} 
\nonumber \\
I_3 &=& \int{d\rho\over \rho(\rho-1)^2} .
\end{eqnarray}
The integral $I_3$ is elementary and is given by
\begin{equation}
I_3=\ln{\rho\over \rho-1}-{1\over \rho-1}.
\end{equation} 

Considering the substitution
\begin{equation}
\rho = b\cosh\theta
\label{rhotheta}
\end{equation} 
the integrals under consideration become
\begin{eqnarray}
I_1 &=& \int d\theta
\left[
{1\over b\cosh\theta}-{1\over b\cosh\theta-1}
+{1\over (b\cosh\theta-1)^2}
\right]
\nonumber \\
I_2&=& \int d\theta b^2(\cosh^2\theta-1)
\left[
{1\over b\cosh\theta}-{1\over b\cosh\theta-1}
+{1\over (b\cosh\theta-1)^2}
\right]
\label{intj}
\end{eqnarray}

In order to obtain the closed form expressions 
of the integrals $I_1$ and $I_2$, we define the two integrals 
$K$ and $L$ as 
\begin{eqnarray}
K &=& \int{d\theta \over \cosh\theta-\epsilon} 
\nonumber \\
L &=& \int{d\theta \over (\cosh\theta-\epsilon)^2} ,
\end{eqnarray}
with $\epsilon=1/b$.

For $\epsilon<1$, making use of the half angle formula
for $\cosh\theta$ and the substitution
\begin{equation}
\tanh{\theta\over 2}=\sqrt{1-\epsilon\over 1+\epsilon} u,
\end{equation} 
we obtain
\begin{equation}
K = {2\over\sqrt{1-\epsilon^2}}
\tan^{-1}\sqrt{1+\epsilon\over 1-\epsilon} \tanh{\theta\over2}.
\end{equation}

On the other hand, for $\epsilon>1$, then the substitution we consider is
\begin{equation}
\tanh{\theta\over 2}=\sqrt{\epsilon-1\over \epsilon+1} v,
\end{equation}
which leads us to
\begin{equation}
K = {2\over\sqrt{\epsilon^2-1}}\int {dv\over v^2-1} =
-{2\over\sqrt{\epsilon^2-1}}\coth^{-1}\sqrt{\epsilon+1\over \epsilon-1}
\tanh{\theta\over2}.
\end{equation}

We also consider the following relation
\begin{equation}
{d\over d\theta}\left({\sinh\theta\over \cosh\theta-\epsilon}\right)
=-{\epsilon\over\cosh\theta-\epsilon}+{1-\epsilon^2
\over(\cosh\theta-\epsilon)^2},
\end{equation}
from which we obtain
\begin{equation}
L = {\epsilon\over 1-\epsilon^2}K+{1\over1-\epsilon^2}{\sinh\theta
\over \cosh\theta-\epsilon}.
\end{equation}

The integrals $I_1, I_2$ in Eq.~(\ref{intj}) can be finally expressed as
\begin{eqnarray}
I_1 &=& {1\over b}\tan^{-1}\sinh\theta-{K\over b}+{L\over b^2}
\nonumber \\
I_2&=&-b\tan^{-1}\sinh\theta +{(b^2+1)\over b}K-{(b^2-1)\over b^2}L.
\end{eqnarray}

Combining these results, for $\epsilon<1$, an explicit expression
of the integral $I$ is given by
\begin{eqnarray}
I &=& \rho_i\cos2\phi\csc\phi\tan^{-1}\sinh\theta
+{\epsilon^2-\csc^2\phi \over 1-\epsilon^2}{\sinh\theta
\over \cosh\theta - \epsilon}
\nonumber \\
&+&
{4\epsilon^2\cot^2\phi-2\epsilon^2+2\csc^2\phi \over 
\epsilon(1-\epsilon^2)^{3/2}}
\tan^{-1}\sqrt{1+\epsilon\over 1-\epsilon}
\tanh{\theta\over2}
\nonumber \\
& \pm&
2\rho_i\cos\phi 
\left(\ln{\rho\over\rho-1}
-{1\over\rho-1}\right).
\label{strom1}
\end{eqnarray}

Also, for $\epsilon>1$, we have
\begin{eqnarray}
I &=& \rho_i\cos2\phi\csc\phi\tan^{-1}\sinh\theta
+{\epsilon^2-\csc^2\phi \over 1-\epsilon^2}{\sinh\theta
\over \cosh\theta - \epsilon}
\nonumber \\
&-&
{4\epsilon^2\cot^2\phi-2\epsilon^2+2\csc^2\phi \over 
\epsilon(\epsilon^2-1)^{3/2}}
\coth^{-1}\sqrt{\epsilon+1\over \epsilon-1}
\tanh{\theta\over2}
\nonumber \\
&-&
2\rho_i\cos\phi 
\left(\ln{\rho\over\rho-1}
-{1\over\rho-1}\right).
\end{eqnarray}

Noting Eq.(\ref{rhotheta}), we may immediately see that Eq.(\ref{strom1})
reduces to Eq.(\ref{eqifront1}).  
In Fig.~\ref{ifront}, we show ionization fronts
for a range of ionizing luminosities characterized by the parameter
$X$. The cool giant is located at the origin and the white dwarf
is at $(\rho_i=80,0)$. We also approximate the
ionization corresponding to $X=0.02$ by a hyperbola
given by $y^2=6.5(x-40)^2-400$.

\end{document}